\documentclass[
reprint,
superscriptaddress,
 amsmath,amssymb,
 aps,
prb,
]{revtex4-2}

\usepackage{graphicx}
\usepackage{dcolumn}
\usepackage{bm}
\usepackage{xcolor}

\usepackage{comment}



\begin{document}

\preprint{APS/123-QED}

\title{Magnetic order and excitations in the magnetically intercalated van der Waals material Cr$_{\frac{1}{4}}$NbSe$_2$}

\author{Ryota Yamaoka}
\author{Hiraku Saito}
\affiliation{The Institute for Solid State Physics, University of Tokyo, Kashiwa, Chiba 277-8581, Japan.}
\author{Yuki Settai}%
\author{Xiang Huang}%
\affiliation{%
Department of Applied Physics, University of Tokyo, Tokyo 113-8654, Japan.
 }%
\author{Daisuke Nishio-Hamane}%
\author{Shingo Takahashi}%
\affiliation{The Institute for Solid State Physics, University of Tokyo, Kashiwa, Chiba 277-8581, Japan.}
\author{Daichi Ueta}%
\affiliation{Institute of Materials Structure Science, High Energy Accelerator Research Organization, Tsukuba, Ibaraki 305-0801, Japan.}
\author{Tatsuro Oda}%
\affiliation{The Institute for Solid State Physics, University of Tokyo, Kashiwa, Chiba 277-8581, Japan.}
\author{Hodaka Kikuchi}
\author{Tao Hong}
\affiliation{Neutron Scattering Division, Oak Ridge National Laboratory, Oak Ridge, TN 37831-6393, U.S.A.}  
\author{Masaki Nakano}%
\affiliation{%
College of Engineering, Shibaura Institute of Technology, Tokyo 135-8548, Japan.
 }%
\author{Shinichiro Seki}
\affiliation{%
Department of Applied Physics, University of Tokyo, Tokyo 113-8654, Japan.
 }%
\affiliation{Research Center for Advanced Science and Technology, University of Tokyo, Komaba, Tokyo 153-8904, Japan.}
\author{Taro Nakajima}
 \email{taro.nakajima@issp.u-tokyo.ac.jp}
\affiliation{The Institute for Solid State Physics, University of Tokyo, Kashiwa, Chiba 277-8581, Japan.}
\affiliation{Institute of Materials Structure Science, High Energy Accelerator Research Organization, Tsukuba, Ibaraki 305-0801, Japan.}
\affiliation{RIKEN Center for Emergent Matter Science (CEMS), Wako 351-0198, Japan.}

\date{\today}

\begin{abstract}
Cr$_{\frac{1}{4}}$NbSe$_2$ is a triangular lattice magnet in which magnetic Cr$^{3+}$ ions are intercalated to form triangular lattices  between NbSe$_2$ van der Waals layers stacked along the c axis. %
By unpolarized and polarized neutron scattering experiments, we have revealed that the magnetic ground state of this system is a 120$^{\circ}$-type antiferromagnetic order characterized by the magnetic propagation wave vector of $q=(\frac{1}{3}, \frac{1}{3}, 0)$. %
We also performed inelastic neutron scattering measurements using co-aligned single crystals, and determined dispersion relations of magnetic excitations at low temperatures. %
Comparing the observed spectra with calculations based on the linear spin-wave theory, we revealed that the out-of-plane ferromagnetic interaction is fairly strong as compared to the in-plane nearest neighbor antiferromagnetic interaction. %
Although the crystal structure of this system is composed of two-dimensional van der Waals layers, the magnetic order has a three dimensional character, which would be attributed to long-range magnetic interactions mediated by conduction electrons. 
\end{abstract}

\maketitle


\section{Introduction}
Van der Waals (vdW) materials with magnetic ions have attracted considerable attention as platforms for exploring two-dimensional magnetism\cite{Park2025vdWreview}.
These systems are often termed vdW magnets, and typically consist of two-dimensional layers of magnetic ions.  %
These layers are weakly coupled with each other by the vdW force, and thus exfoliable down to a monolayer or few layers. %
One of the famous vdW magnets is the transition-metal phosphorus trisulfide family $M$PS$_3$ ($M$ = Fe, Ni, Mn)\cite{Lee2016FePS3,Lancon2016FePS3,Lancon2018NiPS3,Wildes1998MnPS3}, which provides various magnetic properties depending on the magnetic atoms; for instance, Ising, XY, and Heisenberg spin characters are realized in FePS$_3$, NiPS$_3$ and MnPS$_3$, respectively. %
Another examples are CrI$_3$ and Cr$_2$Ge$_2$Te$_6$, which were investigated by Magneto-Optical Kerr Effect (MOKE) measurements, demonstrating that ferromagnetic order persists down to the monolayer or bilayer limit\cite{Huang2017CrI3,Gong2017Cr2Ge2Te6}. %
Furthermore, it was also shown that the magnetic transition temperature of a monolayer of the metallic vdW ferromagnet Fe$_3$GeTe$_2$ can be tuned by gate voltage\cite{Deng2018Fe3GeTe2}. %
There is also a theoretical study on novel magnetoresistivity in Eu-intercalated graphite\cite{Eu_intercalated_graphene_PhysRevX.12.021010}. %
These results have expanded the scope of vdW magnets and their applications. %

One of the most straightforward ways to design vdW magnets is to find materials with two-dimensional layers of magnetic ions coordinated by nonmagnetic ligands, which are the cases for $M$PS$_3$ family. %
Another strategy is to intercalate magnetic ions into nonmagnetic vdW materials such as NbS$_2$, NbSe$_2$, TaS$_2$, and TaSe$_2$, which are known as transition-metal dichalcogenides (TMDs). %
These materials consist of triangular-lattice vdW layers and can accommodate magnetic ions in the vdW gaps. %
The intercalated magnetic ions are regularly arranged to form superstructures depending on the amount of the intercalation. %
Thus, the chemical formulas of the magnetically intercalated TMDs are described as M$_x$TX$_2$\cite{Laar1971MexNbS2} (M=Fe, Cr, Ni, T=Nb, Ta, and X=S, Se). %
Among them, the $x = 1/3$ compounds in which the intercalated magnetic ions form a $\sqrt{3}a\times\sqrt{3}a$ superlattice structure have been extensively investigated since they exhibit nontrivial magnetic orders. %
A combination between the host structure and the superstructure leads to a chiral crystal structure belonging to the space group of $P6_322$. %
This structural chirality induces helimagnetic orders and chiral soliton lattices in Cr$_{1/3}$NbS$_2$\cite{Togawa2012CrNb3S6} and Cr$_{1/3}$TaS$_2$\cite{Obeysekera2021CrTa3S6}. %
It has also been recently shown that a commensurate non-coplanar triple-$Q$ antiferromagnetic order in Co$_{1/3}$TaS$_2$ induces a large spontaneous anomalous Hall effect in zero field\cite{Park2023CoTa3S6,Takagi2023CoTa3S6}. %
Moreover, colinear antiferromagnetic order in Fe$_{1/3}$NbS$_2$\cite{Wu2022PRXFeNbS2} and Ni$_{1/3}$NbS$_2$\cite{An2023NiNb3S6} and ferromagnetic order in Cr$_{1/3}$NbSe$_2$\cite{Gubkin2016Cr13NbSe2} and Fe$_{1/3}$TaS$_2$\cite{Husremovic2022FeTa3S6} were also reported. %
Despite the fact that they are originally derived from two-dimensional vdW materials, these magnetic orders are considered as three-dimensional long-range orders, indicating the presence of substantial interlayer interactions between the magnetic moments. %
In fact, recent first-principles calculations on M$_x$TX$_2$ systems suggested the existence of relatively large interlayer exchange interactions\cite{Hatanaka2023PRB}. %
To confirm this scenario, experimental studies to determine the exchange parameters are necessary. %
However, they have been carried out only a limited number of systems thus far \cite{Park2023CoTa3S6}.
Furthermore, while the magnetic orders of the $x = 1/3$ compounds have been extensively studied as mentioned above, those in the $x = 1/4$ systems were reported for a limited number of compounds\cite{CoTa4Se8_PRB_2024,CrTa4S8_NatCommun_2025}.

In the present study, we focus on Cr$_{1/4}$NbSe$_2$, which is a member of the $x = 1/4$ systems. %
In previous studies, the crystal structure of this compound was studied by X-ray powder diffraction measurements\cite{Voorhoeve1970Intercalation}, and the magnetic susceptibility measurements suggested an antiferromagnetic transition at around 60--70~K\cite{Voorhoeve1971CrNb4Se8_magnetism,Toporova2020CrxNbSe2}. %
However, the detailed refinements of the crystal structure and the determination of the magnetic structure have not been carried out. %
We thus performed unpolarized and polarized neutron scattering measurements on a single crystal of Cr$_{1/4}$NbSe$_2$, determining the crystal structural parameters and the magnetic structure at low temperatures. %
We also performed neutron inelastic scattering measurements using co-aligned single crystals. %
One of the main contributions of the present study to the body of research on magnetically intercalated TMDs is  the experimental and quantitative evaluation of the in-plane and interlayer exchange interactions from the observed magnetic excitation spectra. %
Although a quasi two-dimensional character would be expected from the crystal structures, the dimensionality of the electronic and magnetic properties in these systems need to be discussed from a microscopic point of view. %
The present results provide a quantitative basis for understanding the three-dimensional character of the magnetic orders in the magnetically intercalated TMDs. %

\begin{table*}[t]
\caption{Refined crystallographic parameters for Cr$_{x}$NbSe$_2$ ($x=0.185$) at 2.3 K. %
The occupancy of Cr site is fixed at $4x$ as detailed in the main text. }
\label{NucRefinement_table}
\begin{ruledtabular}
\begin{tabular}{@{}ccccccc@{}}
Atom & Occupancy & $x$ & $y$ & $z$ & $U_{\mathrm{iso}}$ & Site \\ \hline
Cr  & 0.74 (fixed)      & 0           & 0           & 0            & 0.025(2)   & 2a  \\
Nb1 & 1               & 0           & 0           & 1/4          & 0.002(1)   & 2b  \\
Nb2 & 1               & 0.5085(1)   & 0.0169(2)   & 1/4          & 0.0086(9)  & 6h  \\
Se1 & 1               & 1/3         & 2/3         & 0.6118(1)    & 0.0021(9)  & 4f  \\
Se2 & 1               & 0.16761(6)  & 0.3352(1)   & 0.11857(5)   & 0.0051(3)  & 12k \\
\end{tabular}
\end{ruledtabular}
\end{table*}

\section{Experimental details}

\subsection{Sample preparation and characterization}
A polycrystalline sample of Cr$_{1/4}$NbSe$_2$ was prepared from a stoichiometric mixture of Cr, Nb and Se, which were sealed and heated under a vacuum. %
Bulk single crystals were grown by chemical vapor transport (CVT) method using I$_2$ as the transport agent. %
We performed chemical analysis for one of the single crystals using SEM–EDX (JEOL IT-100) under operating conditions of 15 kV accelerating voltage and a probe current of 0.8 nA. %
Data were corrected using the ZAF method, and pure metals of Cr, Nb, and Se were used as standards.  %
The results of the measurements are summarized in Appendix. %
The amount of the intercalation was determined to be $x=0.185 \pm 0.01$. %
We fixed this value in the crystal structure analysis by the neutron diffraction measurements mentioned in the following. %

We also performed magnetization measurements using a commercial SQUID magnetometer, Magnetic Property Measurement System (MPMS, Quantum Design inc.). %
The measurements confirmed the reproducibility of samples grown under identical conditions in terms of both magnetization and transition temperature. %

\subsection{Neutron scattering}
Unpolarized and polarized neutron elastic scattering experiments were carried out at the POlarized Neutron Triple-Axis spectrometer PONTA installed at the 5G beamhole of Japan Research Reactor 3 (JRR-3) \cite{5G_PONTA}. %
As for the unpolarized measurements, the incident neutrons with the energy of $34.05$ meV was obtained by a pyrolytic graphite (PG) monochromator. %
The spectrometer was operated in a two-axis diffraction mode with the horizontal beam collimation of open-80'-80'. %
A single crystal of Cr$_{1/4}$NbSe$_2$ with a mass of 10 mg was loaded in an  Al cell with a small amount of He gas for thermal exchange. %
We performed three experiments for the same crystal with three different horizontal scattering planes, specifically, 
$(H,H,L)$, $(H,0,L)$, and $(H,2H,L)$ planes. %
The sample cell was attached to the cold head of a $^4$He closed-cycle refrigerator and was cooled to the lowest temperature of 2.3 K. %

As for the polarized neutron scattering measurements, the polarized incident neutron beam was obtained by a Heusler (111) monochromator. %
The spectrometer was operated in a triple-axis mode. %
The spin states and the energy of the scattered neutrons were analyzed by a Heusler (111) analyzer. %
The sample was cooled to low temperatures in the same manner as that in the unpolarized measurements. %
The directions of the neutron spins were controlled by a spin flipper and guide fields from a Helmholtz coil. %
The magnetic field at the sample position was approximately 5 mT. %

We also performed neutron inelastic scattering measurements at the high-resolution chopper spectrometer HRC\cite{HRC} in the Materials and Life-science experimental Facility (MLF) of Japan Proton Acceleration Research Complex (J-PARC). %
We used 10 pieces of single crystals (total mass of 1.4 g), which were co-aligned on Al plates with the $(H,H,L)$ horizontal scattering plane. %
The Al plates with the samples were loaded in an Al cell, which was attached to a $^4$He closed-cycle refrigerator. %
The pulsed incident neutron beam with the energy of $E_{i}=15$ meV was obtained by a Fermi chopper rotating with the frequency of
100 Hz. %
The energy resolution at the elastic position was approximately 5.5\%. %
We obtained time-of-flight neutron scattering data for an $\omega$ angle, which is the rotation axis of the sample about the vertical direction, and repeated the measurements with different $\omega$ angles. 
Finally, all the data were processed by MSlice software in DAVE program package\cite{DAVE}. %

To further investigate low-energy magnetic excitations, we carried out additional inelastic neutron scattering measurements on the cold neutron triple-axis spectrometer (CTAX) at the High Flux Isotope Reactor, Oak Ridge National Laboratory. %
The measurements were performed using the same co-aligned crystals as in the HRC experiment. The spectrometer was operated with the fixed final neutron energy of 3.0 meV.  %
The horizontal beam collimation was guide-open-80'-open. %
The incident and final neutron energies were selected by a PG(002) monochromator and a PG(002) analyzer, respectively. %
A cooled Be-filter was placed in front of the analyzer to suppress the high energy neutrons coming from the higher order reflections at the monochromator. %
The energy resolution at the elastic condition was 0.116 meV. 

To examine possible spin fluctuations near the magnetic phase transition temperature, we also performed neutron spin-echo spectroscopy at iNSE spectrometer in JRR-3. %
The incident neutron with the wavelength of 7.3 \AA\ was obtained by a neutron velocity selector. %
The resolution was approximately $d\lambda/\lambda=10\%$. %
Similarly to the CTAX experiment, we used the same co-aligned crystals as that in the HRC experiment. %
A $^4$He closed-cycle refrigerator was used to control the temperature of the sample. %

\section{Results and discussions}

\subsection{Crystal structure}\label{Sec_crystalstr}
\begin{figure}[t]
\includegraphics[keepaspectratio,width=0.95\linewidth]{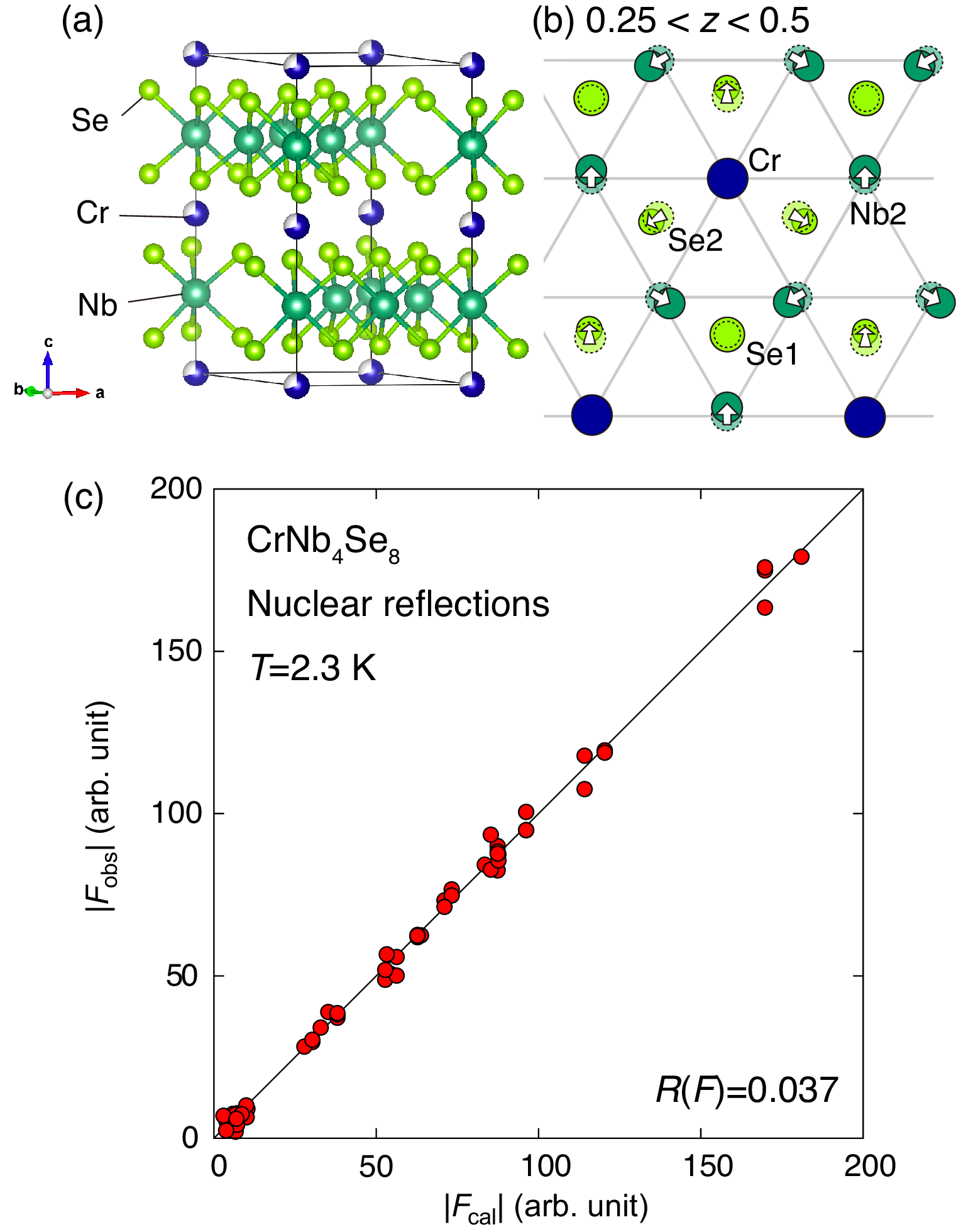}
\caption{\label{CrystStr} (a) Crystal structure of Cr$_{1/4}$NbSe$_2$. %
(b) Schematic illustration showing the atomic displacements near Cr atoms obtained from the structural refinement. %
The white arrows qualitatively show the directions of the displacements. The light-green circles with broken circles inside show the Se1 atoms shifted upward. %
(c) Comparison between the observed and calculated nuclear structure factors, $|F_{\mathrm{obs}}|$ and $|F_{\mathrm{cal}}|$, respectively.}
\end{figure}

Cr$_{1/4}$NbSe$_2$ has a layered structure derived from 2H--NbSe$_2$. %
As shown in Fig. \ref{CrystStr}(a), Nb atoms form triangular lattice layers, and Se atoms coordinate around each Nb atom in a trigonal-prismatic geometry. %
This NbSe$_2$ layer works as the main structural unit of the crystal. %
Cr atoms are intercalated in the vdW gaps between the layers, and forms the $2a\times 2b \times c$ superlattice structure. %
The space group of this compound is $P6_3/mmc$, which preserves inversion symmetry in contrast to the $x=1/3$ systems having the chiral crystal structures. %

In the present unpolarized neutron diffraction measurements, we measured integrated intensities of 65 nuclear Bragg reflections by the $\theta$-$2\theta$ scans, among which 41 reflections are independent, at 2.3 K. %
Each integrated intensity was converted to the structure factor by taking into account the Lorentz factor of $1/\sin 2\theta$\cite{Shirane_TAS}, where $2\theta$ is the scattering angle for the Bragg reflection. %
The effect of neutron absorption was neglected since the absorption cross section of this compound for the incident energy of the present experiment was small enough. %
We refined the atomic coordinates and the isotropic displacement parameters $U_{\mathrm{iso}}$ by the non-linear least-squares method, while fixing the Cr-site occupancy at the value obtained from SEM--EDX measurements. %
The calculated structure factors show good agreement with the observed ones, giving an $R(F)$-factor of 3.7\%, as shown in Fig. \ref{CrystStr}(c).
The refined atomic positions are listed in Table \ref{NucRefinement_table}. %
Note that the site occupancies are calculated using the unit cell of Cr$_{1/4}$NbSe$_2$, which is four-times larger than that of NbSe$_2$. Thus, the occupancy of Cr was fixed at $4x=0.74$. %
The lattice constants were also determined to be $a=b=6.900(3)$ \AA\  and $c=12.52(1)$ \AA. 

From the refinement, it was found that the averaged structure of the NbSe$_2$ layers is nearly the same as that in pristine NbSe$_2$\cite{NbSe2_MEERSCHAUT20011721}. %
This would infer that the system still has two-dimensional character from the viewpoint of crystal structure. %
By looking at the local atomic coordinates near Cr atoms, we found that the Cr intercalation introduce slight displacements of Se and Nb atoms. %
As schematically illustrated in Fig. \ref{CrystStr}(b), the Se2 sites are slightly moved away from the nearest Cr site in both in-plane and out-of-plane directions. %
Accordingly, Nb2 positions shifted in-plane directions compensating the Se2-Nb2 distance. %
Finally, Se1 sites, which are surrounded by Nb2 sites, are slightly moved along the $c$ axis, because the in-plane positions of Se1 are fixed by the site symmetry. %

We note here that the nuclear reflections having odd values of $H$ and/or $K$ indices were clearly observed in the present experiment. %
The odd values of $H$ and $K$ correspond to half-integers of $H$ and $K$ in NbSe$_2$, in which the $a^*$ and $b^*$ vectors are twice as long as those in the $x=1/4$ system. %
The existence of these reflections indicates that the intercalated Cr atoms indeed constitute the $2a\times 2b \times c$ structure despite the imperfect occupancy. %
We observed no additional nuclear superlattice peaks, suggesting that the vacancies of the Cr sites are randomly distributed in the sample. %

It should also be noted here that the pristine NbSe$_2$ exhibit a charge density wave (CDW) order with the modulation wavevector of $(1/3,0,0)$ at low temperatures\cite{NbSe2_CDW_JACS}. %
In the present measurements, we did not observe superlattice reflections indexed by this wavevector. %
One possible interpretation would be that the Cr intercalation suppressed the CDW. %
Another possibility would be that the present neutron measurements, which probe the distribution of the nuclei in the sample, does not have enough sensitivity to detect nuclear displacements induced by the possible CDW order. %
This point needs to be studied using other probes with high sensitivity to the electron distributions, specifically X-ray or electron diffraction measurements, in the future. 

\begin{figure}[t]
\includegraphics[keepaspectratio,width=0.95\linewidth]{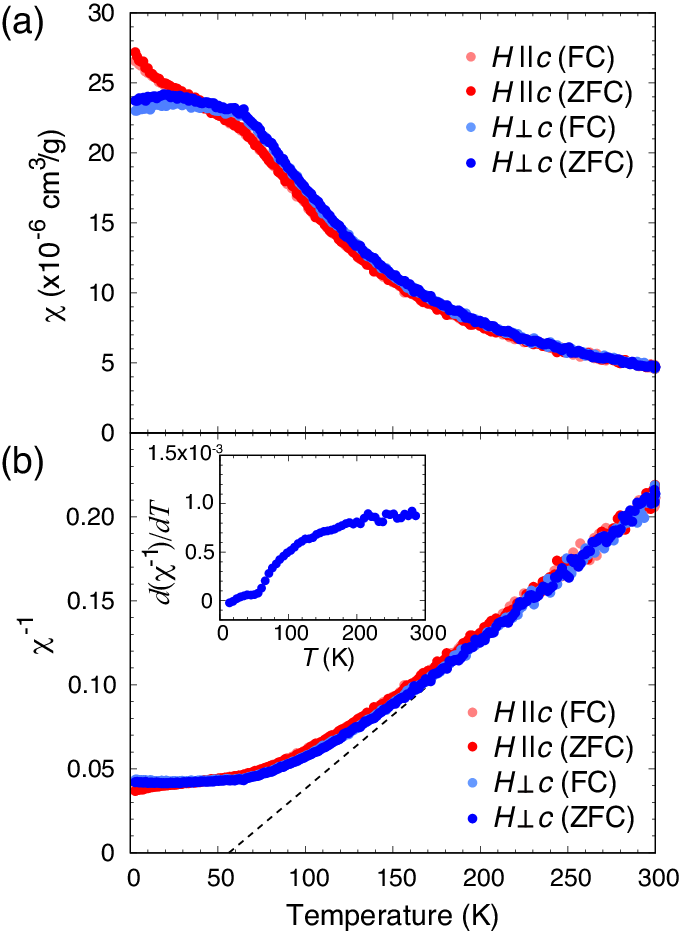}
\caption{\label{Magnetization} (a) Temperature variations of mass-normalized magnetic susceptibilities of Cr$_{1/4}$NbSe$_2$ measured under different field directions. %
All the data were measured on heating under an external field of 100 Oe after FC and ZFC. %
(b) Temperature variations of inverse susceptibilities. The dashed line shows the Curie-Weiss fit for the data in the temperature range of $T>200$ K. %
Inset shows the temperature derivative of the inverse susceptibility for the data labeled "$H\perp c$ (ZFC)".   }
\end{figure}

\begin{figure}[t]
\includegraphics[keepaspectratio,width=0.95\linewidth]{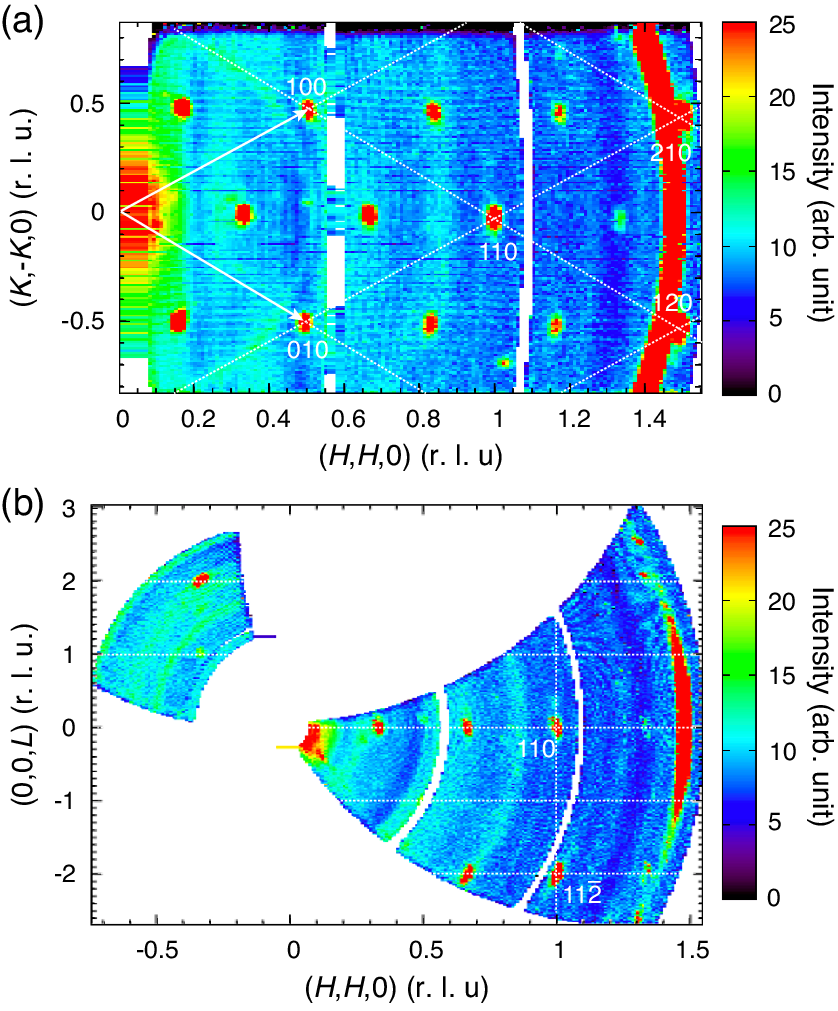}
\caption{\label{IntMaps} [(a),(b)] Magnetic elastic scattering intensity maps for the (a) $(H,K,0)$ and (b) $(0,0,L)$ planes measured at HRC. %
White (dotted) lines are guides to the eyes. }%
\end{figure}

\subsection{Magnetic susceptibility}
We measured temperature variations of magnetic susceptibility, $\chi$, using a single crystal Cr$_{1/4}$NbSe$_2$ sample under an external magnetic field of $100~\mathrm{Oe}$. %
All the data were collected on heating after zero-field cooling (ZFC) and field cooling (FC) for two field directions, $H \parallel c$ and $H \perp c$. %
The results are shown in Fig. \ref{Magnetization}(a).

At high temperatures, the susceptibilities for the two field directions almost overlap with each other. %
This indicates that the magnetic anisotropy in this system is relatively small and that the magnetic moments in this system can be approximated to be Heisenberg spins. %

At low temperatures, a kink anomaly of the susceptibility was observed at around $T_{\rm N}=65$ K, indicating the presence of a magnetic phase transition at this temperature. %
No significant difference between the FC and ZFC curves was found, suggesting the absence of spin-glass-like irreversibility in the measured temperature range.

To estimate an averaged magnetic interactions in this system, the inverse susceptibility $\chi^{-1}(T)$ is plotted in Fig.~2(b). %
The inset shows the temperature derivative of the inverse susceptibility measured under the field perpendicular to the $c$ axis, indicating that it becomes nearly constant above approximately 200 K. %
By fitting the Curie-Weiss (CW) law to the data measured in the temperature range, we obtained $\Theta_{\mathrm{CW}} \sim 60$ K, which indicates, within the mean-field picture, that the exchange interactions in this material is ferromagnetic in average.

We also estimated the effective magnetic moment from the CW fit. %
Taking into account the imperfect Cr-site occupancy, the effective magnetic moment is $\mu_{\mathrm{eff}} \simeq 4.05~\mu_{\mathrm{B}}/\mathrm{Cr}.$
This value is comparable to the spin moment expected for Cr$^{3+}$ ($3d^3$, $S = 3/2$, $g=2$), $\mu_{\mathrm{eff}}^{\mathrm{free}}
= g\sqrt{S(S+1)}= 3.87~\mu_{\mathrm{B}}$.

\begin{figure}[t]
\includegraphics[keepaspectratio,width=0.95\linewidth]{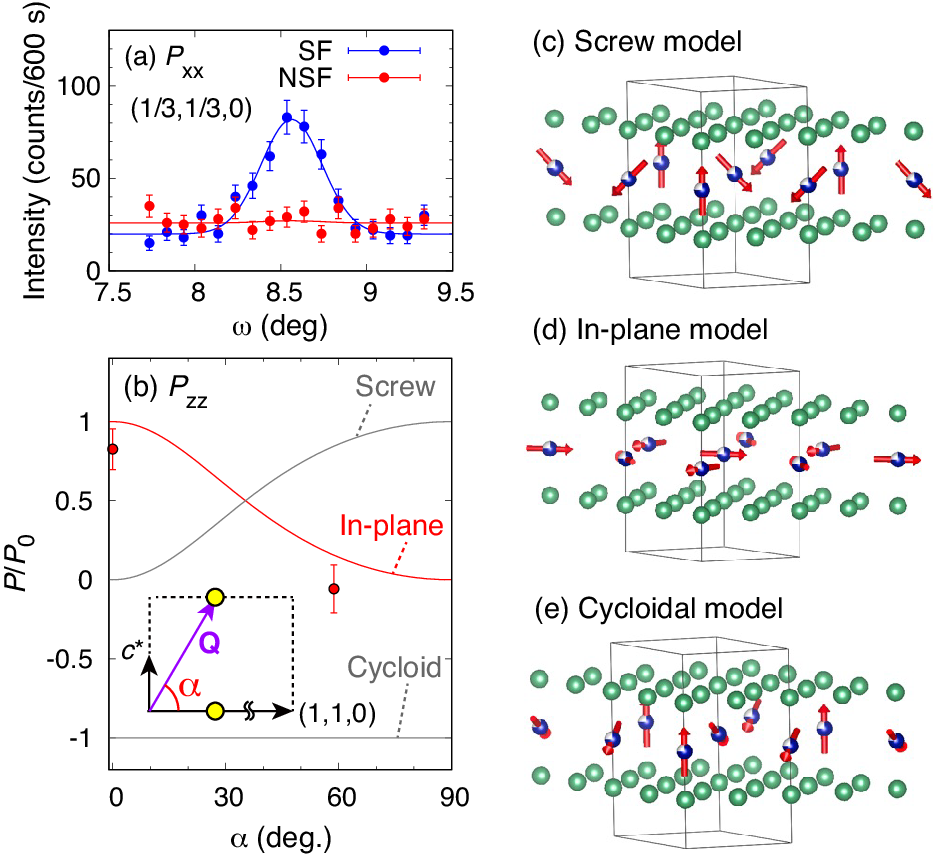}
\caption{\label{PolAnalysis} (a) Scattering profiles of the magnetic Bragg reflections at $(1/3, 1/3, 0)$ measured in the $P_{xx}$ longitudinal polarization analysis. %
(b) Polarization of the scattered neutron beam measured in the $P_{zz}$ longitudinal polarization analysis. %
The data are normalized to the instrument beam polarization $P_0$ and plotted as function of $\alpha$. %
Inset shows the definition of the angle $\alpha$ on the $(H,H,L)$ horizontal scattering plane. %
Yellow circles show the positions of the magnetic Bragg reflections measured in the $P_{zz}$ setup. %
[(c)-(e)] Magnetic structure models characterized by the magnetic modulation wave vector of $\mathbf{q}=(1/3, 1/3,0)$. 
}
\end{figure}

\subsection{Magnetic structure}
Figures \ref{IntMaps}(a) and \ref{IntMaps}(b) show the elastic intensity maps measured at 5 K at HRC. %
We observed nuclear Bragg reflections indexed as $HKL$ where $H$, $K$ and $L$ are all integers. %
In addition, possible magnetic reflections characterized by a wavevector of $\mathbf{q}=(1/3,1/3,0)$ were observed on $L=2n$ planes where $n$ is an integer. %

To examine whether the additional reflections are of magnetic origin, we performed polarized neutron scattering measurements at PONTA. %
In the following, we employ the Cartesian coordinate $xyz$ in which the $x$ and $z$ axes are defined to be parallel to the scattering vector $\mathbf{Q} (=\mathbf{k}_i-\mathbf{k}_f)$ and perpendicular to the scattering plane, respectively. %
Note that $\mathbf{k}_i$ and $\mathbf{k}_f$ are wavevectors for the incident and scattered neutrons, respectively. %
The $y$ axis is defined so as to complete the right-hand coordinate system. %
The spectrometer was first operated in the $P_{xx}$ longitudinal polarization analysis mode, in which the neutron polarization was set to be parallel to the $x$ axis. %
In this mode, the nuclear and magnetic scattering intensities are observed in the non-spin-flip (NSF) and spin-flip (SF) channels, respectively \cite{5G_PONTA}. %
Figure \ref{PolAnalysis}(a) shows the polarized neutron scattering profiles of the reflection at $(1/3,1/3,0)$. %
We found that the intensity was dominated by the SF scattering, revealing that this reflection is a pure magnetic reflection. %

\begin{figure}[t]
\includegraphics[keepaspectratio,width=0.95\linewidth]{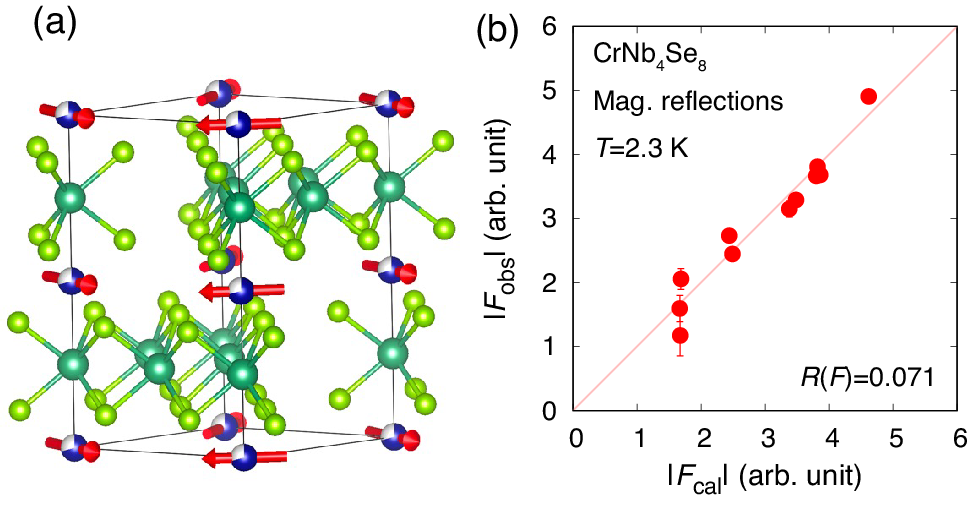}
\caption{\label{MagStr} (a) The in-plane $120^{\circ}$-type magnetic structure of Cr$_{1/4}$NbSe$_2$. %
(b) Comparison between the observed and calculated magnetic structure factors, $|F_{\mathrm{obs}}|$ and $|F_{\mathrm{cal}}|$, respectively.}
\end{figure}

The magnetic reflection with the $q$-vector of $(1/3,1/3,0)$ implies that the magnetic structure is a 120$^{\circ}$-type antiferomagnetic order. %
Similarly to other triangular lattice antiferromagnets showing 120$^{\circ}$-type orders, the magnetic moments are expected to be in the $ab$ plane. %
However, the $P_{xx}$ measurements do not provide any information on the direction of the magnetic moments. %
We thus performed $P_{zz}$ longitudinal polarization analysis measurements, in which the neutron polarization was set to be parallel to the $z$ axis. %
Recalling that a magnetic scattering intensity is proportional to an absolute square of Fourier-transformed magnetic moments projected onto the plane perpendicular to the scattering vector, $|\mathbf{M}^{\perp}(\mathbf{Q})|^2$, $P_{zz}$ configuration enables us to separate the $y$ and $z$ components of $\mathbf{M}^{\perp}(\mathbf{Q})$ into the SF and NSF intensities, respectively. %
Thus, the polarization of neutrons scattered by a magnetic Bragg reflection is written as follows: 
\begin{eqnarray}
P=\frac{I_{\rm NSF}-I_{\rm SF}}{I_{\rm NSF}+I_{\rm SF}}=\frac{|M^{\perp}_{z}(\mathbf{Q})|^{2}-|M^{\perp}_{y}(\mathbf{Q})|^{2}}{|M^{\perp}_{z}(\mathbf{Q})|^{2}+|M^{\perp}_{y}(\mathbf{Q})|^{2}},
\end{eqnarray}
where $I_{\rm NSF}$ and $I_{\rm SF}$ are the integrated intensities of the magnetic Bragg reflection measured in the NSF and SF channels, respectively. %
We performed the $P_{zz}$ polarization analysis for the magnetic Bragg reflections at $(1/3,1/3,0)$ and $(1/3,1/3,2)$. %
Figure \ref{PolAnalysis}(b) shows the polarization of the observed magnetic reflection as function of $\alpha$, which is an angle between the $(H,H,0)$ direction and the $\mathbf{Q}$-vector when measuring the magnetic reflection. %
The data are normalized to the instrument beam polarization $P_0$, which was determined to be 0.877 by measuring nuclear Bragg reflections, in order to correct the effect of the imperfect beam polarization. %
We also show calculated curves for the possible 120$^{\circ}$-type structures with different orientations of the magnetic moments. %
The experimental data agree with the model in which the Cr magnetic moments lie in the $ab$ plane.
We should recall here that the magnetic reflections were observed only on $L=$even planes, indicating that the magnetic modulations on the two Cr layers in the unit cell are ferromagnetically coupled. %
From these results, we obtained the model for the magnetic structure as shown in Fig. \ref{MagStr}(a). %

\begin{figure*}[t]
\includegraphics[keepaspectratio,width=0.98\linewidth]{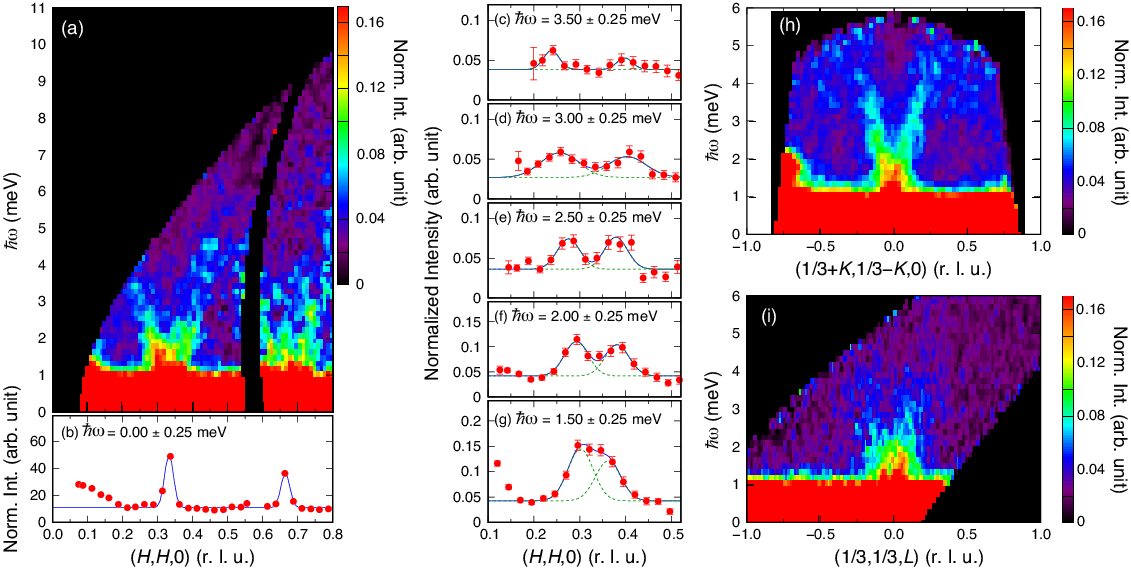}
\caption{\label{HRCdata} (a) Magnetic excitation spectrum along the $(H,H,0)$ in Cr$_{1/4}$NbSe$_2$ measured at 5 K. %
[(b)-(g)]  Constant-energy cuts of the excitation spectra along the $(H,H,0)$ line at different energies. Dotted and solid lines are Gaussian functions fitted to the data. %
[(h),(i)] Magnetic excitation spectra along the (h) $(1/3+K,1/3-K,0)$ and (i) $(0,0,L)$ directions. %
}
\end{figure*}

Finally, we performed unpolarized neutron diffraction measurements at 2.3 K at PONTA to determine the magnitude of the ordered magnetic moments. %
We measured integrated intensities of the magnetic Bragg reflections by $\theta$-$2\theta$ scans \cite{Shirane_TAS}, and converted them into the structure factors following the same procedure as in the crystal structure analysis. %
Figure \ref{MagStr}(b) shows a comparison between the observed and calculated magnetic structure factors. %
We obtained a reasonably good $R(F)$ factor of 7.1\%. 
Taking into account the imperfect Cr-site occupancy, the ordered moment was estimated to be
$1.96(2)~\mu_{\mathrm{B}}/\mathrm{Cr}.$
This value is significantly smaller than the full ordered magnetic moment expected from the results of the magnetic susceptibility measurements, namely $S=3/2$ $(3 \mu_{\rm B})$.
A similar discrepancy between the ordered and effective moments has been reported in Co$_{1/3}$TaS$_2$\cite{Park2023CoTa3S6}, and may be attributed to screening effects by itinerant electrons in the host metallic layers.

\subsection{Magnetic excitations}

To investigate the magnetic interactions in Cr$_{1/4}$NbSe$_2$, we measured magnetic excitations at HRC in MLF of J-PARC.  %
Figure \ref{HRCdata}(a) shows the observed spectrum along the $(H,H,0)$ direction. %
We observed dispersive spin-wave excitations emerging from the elastic magnetic Bragg peaks shown in Fig. \ref{HRCdata}(b). %
Figures \ref{HRCdata}(c)-\ref{HRCdata}(g) show the constant-energy cuts near $H=1/3$, revealing that the two peaks corresponding to the magnon excitations are gradually moving away from each other as the energy transfer $\hbar\omega$ is increased. %
To extract the peak positions, we fitted two Gaussian functions to the data with a constraint that the widths of the Gaussians were the same. %
Similar spin-wave dispersion was also observed in the excitation spectrum for the $(1/3+K,1/3-K,0)$ direction, as shown in Fig. \ref{HRCdata}(h). %
Importantly, we also observed a steep spin-wave dispersion along the $(1/3,1/3,L)$ direction, as shown in Fig. \ref{HRCdata}(i). %
This clearly demonstrates that this system has sizable magnetic interactions along the $c$ axis. %
The constant-energy cuts for the intensity maps in Figs. \ref{HRCdata}(h) and \ref{HRCdata}(i) are shown in Appendix. %

We note here that these inelastic scattering signals were rather weak and broad, and that the signal-to-noise ratio of the present experiment was not high enough to observe the signals in the high energy regions. %
This is partly because the density of the magnetic moments is relatively low in this compound. %
Another reason could be the effect of randomness in the magnetic Cr sites. %
As we discussed in Sec. \ref{Sec_crystalstr}, the triangular lattice layers of the intercalated Cr atoms contain 26\% of atomic vacancies. %
The randomly distributed  vacancies would affect coherent propagations of the spin waves, leading to the broadening of the excitation spectra. %
We also mention that the previous study on Co$_{1/3}$TaS$_2$\cite{Park2023CoTa3S6} also presented neutron inelastic scattering spectra, in which the inelastic scattering signals also seem to be smeared in high energy regions. %

\begin{figure}[t]
\includegraphics[keepaspectratio,width=0.95\linewidth]{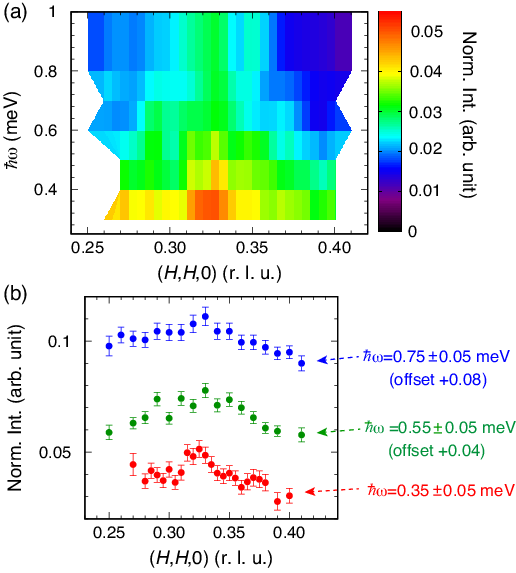}
\caption{\label{HFIRdata} (a) The intensity map of the magnetic excitation spectra measured along the $(H,H,0)$ direction at 1.5 K at CTAX spectrometer in HFIR. %
(b) The constant-energy-transfer profiles extracted from the intensity map shown in (a). %
To enhance the visibility, we summed up the intensities measured at 0.3 and 0.4 meV. %
Similarly, the intensities at 0.5 and 0.6 meV (0.7 and 0.8 meV) were also summed up. %
Each profile is shifted by 0.04 along the vertical direction. }
\end{figure}

To observe the low-energy part of the inelastic scattering spectra below 1 meV, which could not be resolved by the measurements with $E_i=15$ meV at HRC, we performed additional neutron inelastic scattering measurements at CTAX in HFIR. %
Figure \ref{HFIRdata}(a) shows the magnetic excitation spectra measured along the $(H,H,0)$ line at 1.5 K. %
In Fig. \ref{HFIRdata}(b), we also show constant-energy-transfer cuts replotted from the data shown in Fig. \ref{HFIRdata}(a). %
The spin-wave excitations are emerging from the magnetic elastic Bragg peak at $(1/3, 1/3, 0)$. 
We did not observe an energy gap in the spin wave spectra within the energy resolution of this experiment, which is consistent with the Nambu-Goldstone mode expected in a classical Heisenberg model for a triangular lattice antiferromagnet\cite{TLA_SW_PRB2009}.  %

\begin{figure}[t]
\includegraphics[keepaspectratio,width=0.98\linewidth]{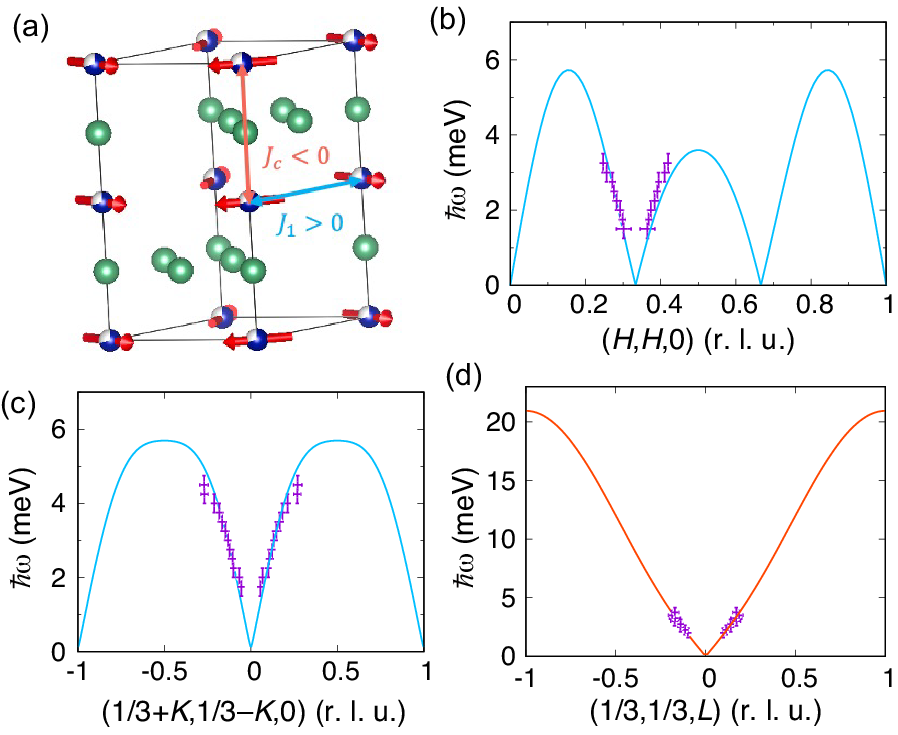}
\caption{\label{SWfitting} (a) The definition of the intra-layer and inter-layer magnetic interactions $J_1$ and $J_c$, respectively. %
[(b)-(c)] Calculated spin-wave dispersion relations fitted to the observed data points for the (b) $(H,H,0)$, (c) $(1/3+K,1/3-K,0)$ and (d) $(1/3,1/3,L)$ lines. %
We sliced the intensity maps with finite widths along the energy-transfer direction, and then extracted the peak positions by the fitting analysis described in the main text. The vertical error bars represent the width of the slicing. The horizontal error bars show the asymptotic standard error of the fitting analysis. %
}
\end{figure}

To analyze the observed data, we employed the minimal Heisenberg model which includes only the in-plane nearest-neighbor (NN) interaction $J_{1}$ and the interlayer nearest-neighbor interaction $J_{c}$, as shown in Fig. \ref{SWfitting}(a). %
The Hamiltonian is written as

\begin{align}
\mathcal{H}
= \sum_{\langle i,j\rangle_{\mathrm{intra}}} J_{1}\,\mathbf{S}_{i}\cdot\mathbf{S}_{j}
+ \sum_{\langle i,j\rangle_{\mathrm{inter}}} J_{c}\,\mathbf{S}_{i}\cdot\mathbf{S}_{j},
\end{align}
where $\langle i,j\rangle_{\mathrm{intra}}$ and $\langle i,j\rangle_{\mathrm{inter}}$ stand for intra- and inter-layer nearest-neighbor spin pairs, respectively. %

By applying the linear spin-wave approximation to this Hamiltonian and the 120$^{\circ}$-type antiferromagnetic ground state, the excitation energy is written as follows.

\begin{align}
\hbar\omega_{\mathbf{q}}
= S \Biggl[
& \left\{
   J_{1}\!\left(3 + 2\gamma_{\mathbf{q}}^{(1)}\right)
 + 2J_{c}\!\left(-1 + \gamma_{\mathbf{q}}^{(2)}\right)
  \right\}
\nonumber\\
&\times
  \left\{
   J_{1}\!\left(3 - \gamma_{\mathbf{q}}^{(1)}\right)
 + 2J_{c}\!\left(-1 + \gamma_{\mathbf{q}}^{(2)}\right)
  \right\}
\Biggr]^{1/2},
\end{align}

where

\begin{align}
\gamma_{\mathbf{q}}^{(1)} &= \cos 2\pi H + \cos 2\pi K + \cos 2\pi(H+K), \\
\gamma_{\mathbf{q}}^{(2)} &= \cos \pi L.
\end{align}

We fitted the calculated dispersions to the peak positions observed along the $(H,H,0)$, $(1/3+K,1/3-K,0)$ and $(1/3,1/3,L)$ lines, as shown in Figs. \ref{SWfitting}(b)-\ref{SWfitting}(d). %
As a result, the exchange interactions were estimated to be $J_{1}S = 1.80 \pm 0.05~\mathrm{meV}$ and $J_{c}S = -4.30 \pm 0.64~\mathrm{meV}$. %
These values qualitatively agree with the results of the magnetic susceptibility measurements which pointed out the existence of relatively large ferromagnetic interactions in this system. %
Although the crystal structure of this system consists of the two-dimensional vdW layers, the present results show that the magnetic Cr ions separated by the layers are strongly coupled with each other by long-range interactions. %
In fact, the previous $ab$-$intio$ calculation\cite{Hatanaka2023PRB} predicated that the inter-layer magnetic interaction is stronger than the intra-layer interaction in Cr$_{1/4}$NbSe$_2$. %
The $ab$-$intio$ calculation also presented the magnetic interactions for the $x=1/3$ and $1/4$ systems with various  intercalated ions and host-TMD layers. %
As for the Cr-intercalated $x=1/4$ systems, the inter-layer and intra-layer magnetic interactions do not largely depend on the host-TMD layers\cite{Hatanaka2023PRB}. %
This is consistent with the fact that the ferromagnetically-stacked 120$^{\circ}$-type order was also reported in Cr$_{1/4}$TaS$_2$\cite{CrTa4S8_NatCommun_2025}. %
By changing the intercalated ions from Cr to Co while fixing the amount of the intercalation at $x=1/4$, the calculated inter-layer interaction turns into antiferromagnetic regardless of the host-TMD layers, while the intra-layer interaction is predicted to change the sign depending on the host-TMD layers\cite{Hatanaka2023PRB}. %
This also agrees with the $A$-type antiferromagnetic order reported for Co$_{1/4}$TaSe$_2$\cite{CoTa4Se8_PRB_2024}. %

We note that the energy range of the observed peak positions along the $(1/3, 1/3, L)$ line is rather limited as compared to the energy scale of the dispersion relation predicted by the calculation. %
Although the observed steep spin-wave dispersion along the $L$ direction unequivocally showed the existence of the strong magnetic interaction along the $c$ axis, the magnitude of $J_c$ would be further refined by examining other models with distant interactions and by measuring the spectra in the high energy region in more detail in the future. %

Let us compare the present results with the other vdW magnets reported in the previous studies. %
Table~\ref{tab:JcJab_comparison} summarizes the ratios of the inter-layer magnetic interactions to the strongest in-plane interactions, which are referred to as $J_c$ and $J_1$ in this paper, respectively, for representative vdW magnets and related layered magnets. %
Although some of them were analyzed by assuming the existence of the second- or third-nearest-neighbor in-plane interactions, we show only the strongest interaction in the table for simplicity. %
In typical vdW magnets such as $M$PS$_3$ 
\cite{Lancon2016FePS3,Lancon2018NiPS3,Wildes1998MnPS3}, CrPS$_4$ \cite{Calder2020CrPS4} 
and $M$X$_3$
\cite{Chen_PhysRevX.11.031047,Schneeloch2022CrCl3,CrBr3_PhysRevB.104.L020402},
the ratio is less than 10\% or assumed to be zero, meaning that the two-dimensional magnetic layers are nearly isolated from each other. %
This agrees with the original expectation for the vdW magnets. %
As for the layered antiferromagnetic topological insulator %
MnBi$_2$Te$_4$ \cite{Li2020MnBi2Te4}, %
the ratio is approximately ${\sim}20$\% and thus relatively larger than those for the vdW magnets mentioned above.
By contrast, magnetically intercalated TMDs, such as Co$_{1/3}$TaS$_2$ \cite{Park2023CoTa3S6} and the presently studied Cr$_{1/4}$NbSe$_2$, exhibit remarkably large ratios of 75--240\%, highlighting the strong interlayer magnetic coupling. %

\begin{table*}[t]
\caption{\label{tab:JcJab_comparison}
Comparison of the ratio $|J_c|/|J_{1}|$ for representative van der Waals magnets.
All exchange parameters were taken from the previous inelastic neutron scattering studies cited in the Ref. column.}
\renewcommand{\arraystretch}{1.3}
\begin{ruledtabular}
\begin{tabular}{llcccc}
Family & Material & $J_{1}$ (meV) & $J_{c}$ (meV) & $|J_c|/|J_{1}|$ (\%) & Ref. \\
\hline
$M$PS$_3$ & FePS$_3$  & 1.46    & $-0.0073$ & 0.50  & \cite{Lancon2016FePS3} \\
          & NiPS$_3$  & 1.9     & 0         & 0     & \cite{Lancon2018NiPS3} \\
          & MnPS$_3$  & $-0.77$ & 0.0019    & 0.25  & \cite{Wildes1998MnPS3} \\
          & CoPS$_3$  & $-1.37$ & 0         & 0     & \cite{Wildes2023CoPS3} \\
\hline
$M$X$_3$  & CrCl$_3$  & $-0.94$ & 0         & 0     & \cite{Schneeloch2022CrCl3} \\
          & CrI$_3$   & $-2.11$ & 0.048     & 2.3   & \cite{Chen_PhysRevX.11.031047} \\
          & CrBr$_3$   & $-1.48 $ & 0     & 0   & \cite{CrBr3_PhysRevB.104.L020402} \\
\hline
          & CrPS$_4$  & $-2.96$ & 0.16     & 5.4   & \cite{Calder2020CrPS4} \\
          & MnBi$_2$Te$_4$\footnotemark[1] & $-0.12$ & 0.022 & 18.3 & \cite{Li2020MnBi2Te4} \\
\hline
M$_x$TX$_2$     & Co$_{1/3}$TaS$_2$\footnotemark[2] & 1.74 & 1.31 & 75.3  & \cite{Park2023CoTa3S6} \\
          & Cr$_{1/4}$NbSe$_2$\footnotemark[3]     & 1.2   & $-2.87$ & 238.9 & [This work] \\
\end{tabular}
\end{ruledtabular}
\footnotetext[1]{$J_{1}$ and $J_c$ are given as $JS$ in the reference. The above values were obtained by assuming $S=5/2$.}
\footnotetext[2]{$J_{1}$ and $J_c$ are given as $JS^{2}$ in the reference. The above values were obtained by assuming $S=3/2$. }
\footnotetext[3]{$J_{1}$ and $J_c$ are given as $JS$ in the present analysis. The above values were obtained by assuming $S=3/2$. }
\end{table*}

We suggest that these systems would be broadly classified into two groups according to the presence or absence of conduction electrons. %
Specifically, $M$PS$_3$ and $M$X$_3$ are insulators, in which magnetic interactions are often mediated by short-range direct exchange or super exchange interactions. %
On the other hand, Co$_{1/3}$TaS$_2$ and Cr$_{1/4}$NbSe$_2$ are metals, in which couplings between conduction electrons and local magnetic moments would yield long-range magnetic interactions. %
It is also worth mentioning that a recent ARPES study on Cr$_{1/3}$NbSe$_2$ observed dispersive electron bands along the $c$ axis\cite{Bruno_PRR_2022}. %
These results imply that Ruderman-Kittel-Kasuya-Yoshida (RKKY)-type long-range interactions account for the inter-layer coupling in the magnetically-intercalated TMDs. %
To verify this scenario, it would be necessary to carry out further systematic studies on magnetic interactions in  M$_x$TX$_2$ systems. %

\subsection{Temperature dependence of the magnetic correlations}

\begin{figure}[t]
\includegraphics[keepaspectratio,width=0.95\linewidth]{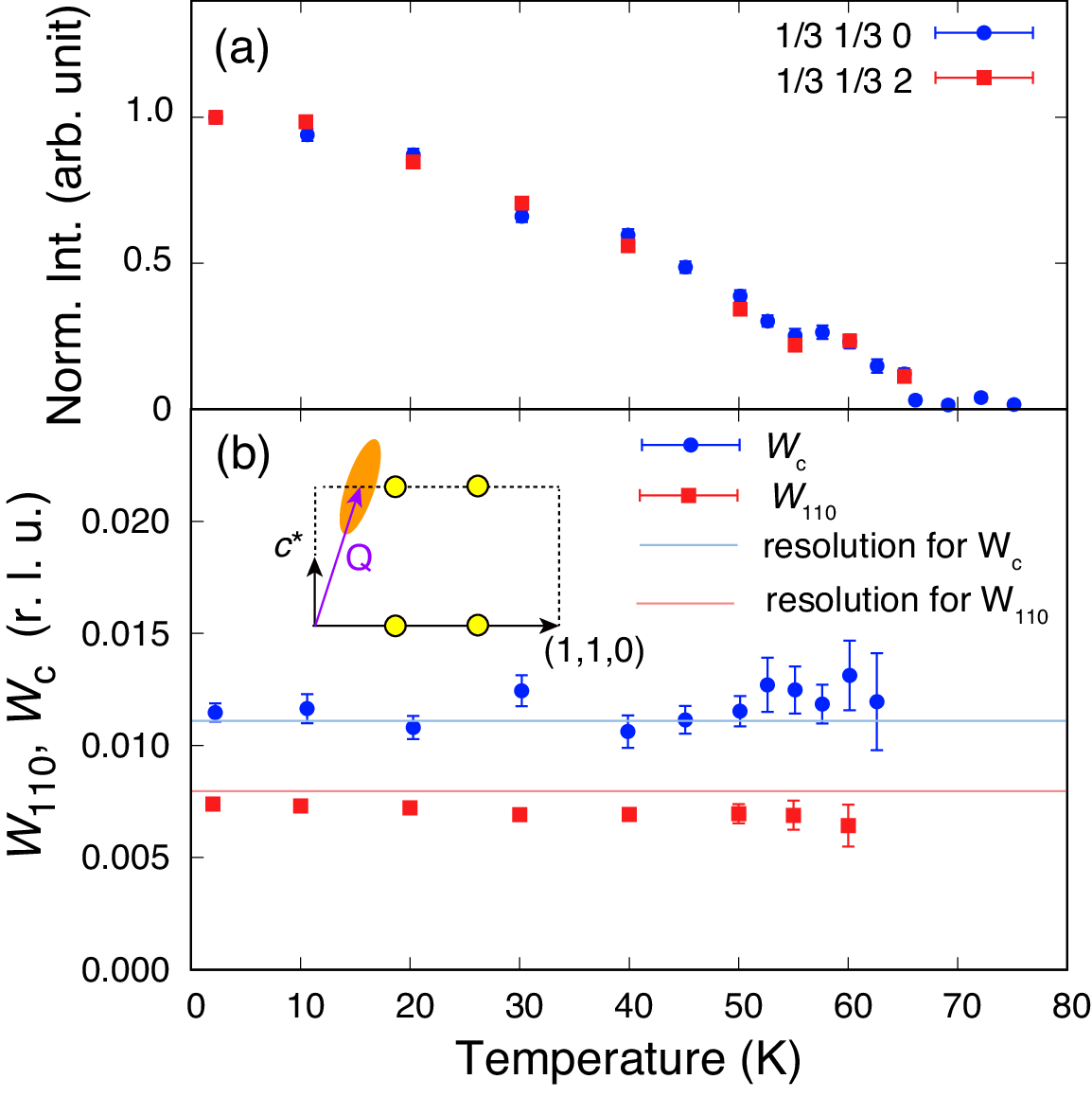}
\caption{\label{Tdep} (a) Temperature dependence of the integrated intensities of the magnetic Bragg reflections at $(1/3,1/3,0)$ and $(1/3,1/3,2)$. These data are normalized to their lowest-temperature values. %
(b) Temperature dependence of $W_{110}$ and $W_{c}$. Solid lines show the resolution widths. %
Inset is a schematic drawing for the resolution function of a triple-axis neutron spectrometer. Orange ellipse qualitatively shows the shape of the resolution functions, and yellow circles denote the positions of magnetic Bragg reflections. %
}
\end{figure}

Figure \ref{Tdep}(a) shows temperature variations of the integrated intensities of the magnetic Bragg reflections at $(1/3, 1/3, 0)$ and $(1/3, 1/3, 2)$ at PONTA in JRR-3. %
These intensities are normalized to those at the lowest temperature. %
They monotonically decrease and nearly coincide with each other at all the temperatures below $T_{N}$. %
This suggests that the 120$^\circ$-type magnetic structure is maintained up to temperatures near  $T_{N}$. %
We also measured the widths of these magnetic Bragg peaks, which are inversely proportional to the magnetic correlation lengths, with varying temperature. %
The widths along the (1,1,0) and $c^*$ directions ($W_{110}$ and $W_c$) were measured by the $(H, H, 2)$ and $(1/3, 1/3, L)$ scans for the magnetic Bragg reflections at (1/3, 1/3, 2) and (1/3, 1/3, 0), respectively. %
The former and the latter widths are inversely proportional to the spin correlation lengths along the (1,1,0) and $c^*$ directions, respectively. %
We note here that the resolution function of a triple-axis neutron spectrometer has an elliptic shape shown in the inset of Fig. \ref{Tdep}(b) at an elastic condition. %
Specifically, the long and short axes of the ellipse are parallel and perpendicular to the scattering vector \cite{Shirane_TAS}. %
Consequently, for the reflection at (1/3,1/3,2), the $(H,H,2)$ scan has a better resolution to detect the possible peak broadening as compared to the scan along the $c^*$ direction. %
For the same reason, the width along the $c^*$ axis was measured at (1/3, 1/3, 0). 
Figure \ref{Tdep}(b) displays the temperature variations of $W_{110}$ and $W_{c}$, which are defined as the full widths at half maxima of the observed scan profiles of the magnetic reflections. %
For both directions, the widths remain comparable to the resolution limits estimated by measuring nuclear Bragg reflections, and no appreciable broadenings were observed even up to temperatures close to $T_{\mathrm{N}}$. %
This demonstrates that the magnetic order retains long-range correlations in both the in-plane and out-of-plane directions even near $T_{\mathrm{N}}$.
This is in contrast to the case of the two-dimensional triangular-lattice antiferromagnet NiGa$_2$S$_4$, in which the magnetic correlation length decreases significantly as the temperature is increased.\cite{Stock2010NiGa2S4} %

\begin{figure}[t]
\includegraphics[keepaspectratio,width=0.95\linewidth]{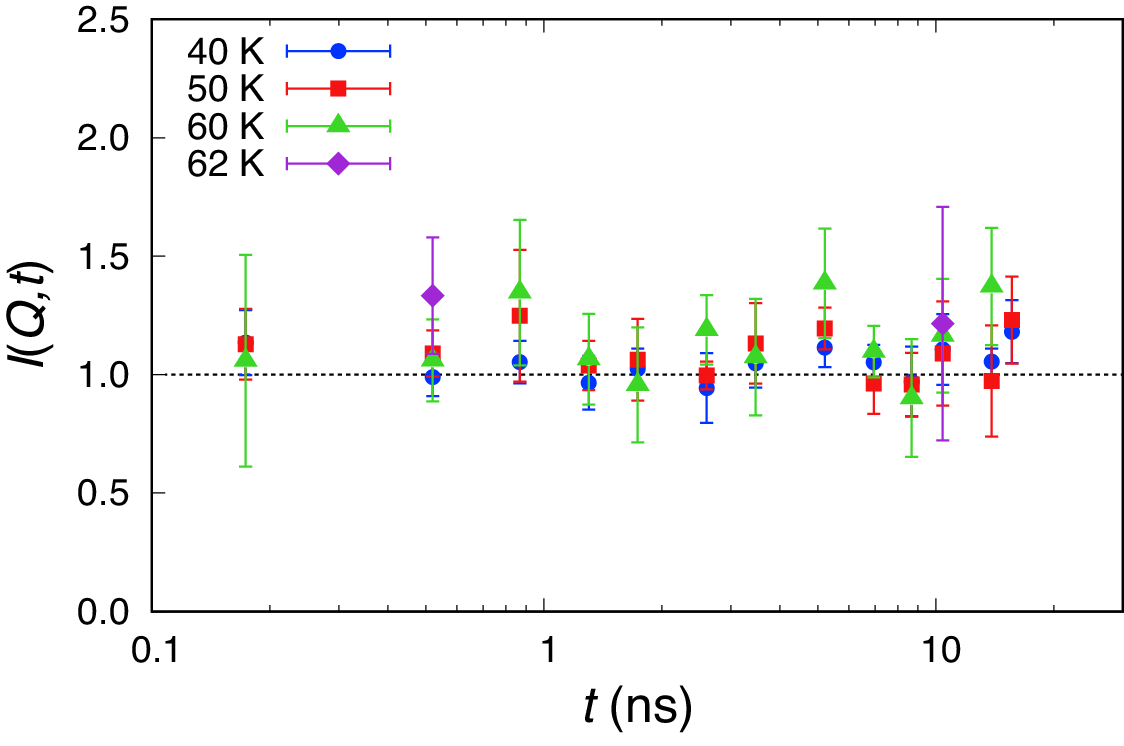}
\caption{\label{NSE} $I({\bf Q},t)$ profiles measured by neutron spin-echo measurements at 40, 50, 60 and 62 K. %
These values are normalized to the $I({\bf Q},t)$ curve measured at the lowest temperature of 2.2 K. }
\end{figure}

We also performed neutron spin-echo measurements at iNSE in JRR-3. %
We focused on the signals near the magnetic Bragg peak of $(1/3,1/3,0)$ and measured them with varying temperature. %
Figure \ref{NSE} shows the intermediate scattering function $I({\bf Q},t)$ at ${\bf Q}=(1/3,1/3,0)$ normalized to that at the lowest temperature of 2.2 K. %
As for the magnetic scattering, this quantity shows how the spin correlation characterized by the wavevector ${\bf Q}$ changes with time $t$. %
In the limit of $t\rightarrow 0$, this quantity corresponds to the instantaneous spin correlation normalized to that at 2 K, and approaches unity. %
If the intensity at ${\bf Q}$ originates from slow spin fluctuations, $I({\bf Q},t)$ decays to zero with increasing $t$. %
In Cr$_{1/4}$NbSe$_2$, the intensity of the magnetic Bragg peak becomes weaker as the temperature is increased. %
By contrast, the $I({\bf Q},t)$ curve does not show a discernible decay within the spin-echo time range of up to 15 ns as shown in Fig. \ref{NSE}. %
This means that the Fourier component at around $(1/3,1/3,0)$ corresponds to a static order. %
It should be noted that the reduction of $I({\bf Q},t)$, which is indicative of spin fluctuations, was sometimes observed in two-dimensional frustrated magnets such as NiGa$_2$S$_4$\cite{Nambu_NiGaS_PRL2015} and CuCrO$_2$\cite{Ehlers_CuCrO2_2013}. %
Although this kind of spin fluctuations should exist even in non-frustrated systems in principle, the above examples demonstrated that they could be enhanced by the low-dimensionality and/or magnetic frustration and that they are detectable by neutron spin echo measurements. %
The present results show that Cr$_{1/4}$NbSe$_2$ does not have such a strong magnetic fluctuations near $T_{\rm N}$, which is distinct from the other systems having strong two-dimensionality.

\section{Summary}
In this study, we have investigated the magnetic order and excitations of the magnetically-intercalated vdW compound Cr$_{1/4}$NbSe$_2$ by neutron scattering measurements. %
As for the crystal structure, the present single-crystal neutron diffraction measurements revealed that, despite the intercalation of Cr atoms, the NbSe$_2$ layers retain their structure in average, while the atoms near the intercalated Cr show slight displacements. %
The polarized and unpolarized neutron scattering measurements demonstrated that the magnetic structure at low temperatures is the in-plane $120^\circ$-type antiferromagnetic structure with magnetic moment lying in the $ab$ plane. %
These magnetic moments are stacked ferromagnetically along the $c$ axis. %
The inelastic neutron scattering measurements revealed steep spin-wave dispersions emerging from the magnetic Bragg peaks.  %
By fitting the calculated dispersion relations, which were derived by applying the linear spin-wave approximation to the model Hamiltonian, to the observed data, we estimated the exchange constants as $J_{1} = 1.80 \pm 0.05~\mathrm{meV}, J_{c} = -4.30 \pm 0.64~\mathrm{meV}.$
The fairly large ferromagnetic interaction along the $c$ axis shows that this system exhibits the three-dimensional magnetic correlation, which is also consistent with the recent theoretical prediction\cite{Hatanaka2023PRB}. %
The measurements on temperature dependence of the magnetic correlation lengths and temporal spin correlations further confirmed this scenario. %
We propose that the coupling between the conduction electrons and local magnetic moments could be the origin of the strong interlayer magnetic interaction in this system. %
This work provides insights into both the static and dynamical properties of the magnetism in Cr$_{1/4}$NbSe$_2$, and contributes to the understanding of the mechanisms by which three-dimensional magnetic order can emerge in van der Waals magnetic materials. %

\section*{Acknowledgements}
The authors are grateful to T. Hatanaka, T. Nomoto, R. Arita and K. Ishizaka for fruitful discussions. %
The neutron scattering experiments at PONTA and the neutron spin-echo measurements at iNSE in JRR-3 were carried out along the proposals (No. 24401 and 25405, respectively) and partly supported by the institute for solid state physics of the university of Tokyo. %
The neutron inelastic scattering experiments at HRC in the Materials and Life Science Experimental Facility of the J-PARC was performed under a user program (Proposal No. 2025S01). %
A portion of this research used resources at the High Flux Isotope Reactor, a DOE Office of Science User Facility operated by the Oak Ridge National Laboratory. The beam time was allocated to CTAX on proposal number IPTS-35059.
The neutron scattering measurement at CTAX in HFIR was also supported by the US-Japan cooperative research program on neutron scattering (Proposal No. 2025-9). %
This work was supported by JSPS KAKENHI (Grant Nos. 21H04990, 22H01949, 22H04965, 24H02235, 25H00611, 25H00846), JST CREST (Grant No. JPMJCR23O4), JST FOREST (Grant No. JPMJFR223W), Murata Science Foundation, Asahi Glass Foundation. %
The part of this study was carried out by the joint research of the Cryogenic Research Center, the University of Tokyo.
S. T.  was supported by JSPS KAKENHI (Grant No. 25KJ0050). %
M. N. was partly supported by The Mitsubishi Foundation and by SIT Supporting Program for Innovative Research (S-SPIRE). %
The images of the crystal and magnetic structures in this paper were depicted using the software VESTA\cite{VESTA}.

\begin{figure}[b]
\includegraphics[keepaspectratio,width=0.95\linewidth]{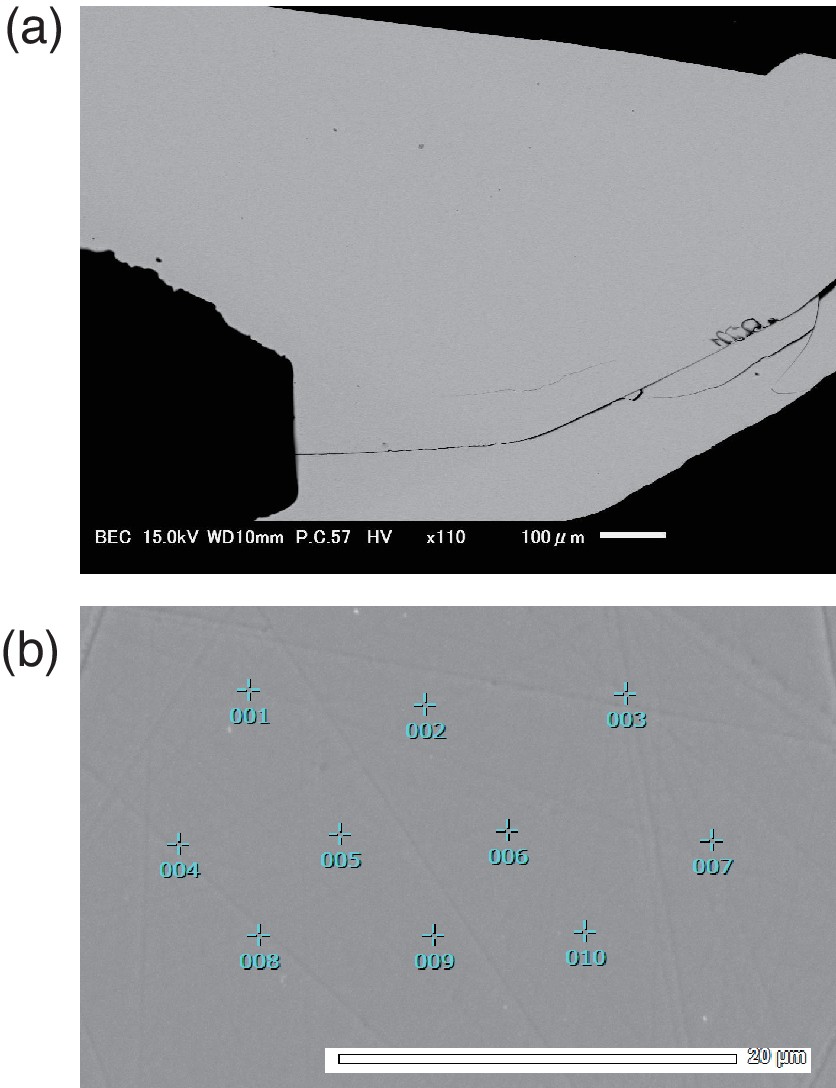}
\caption{\label{SEM} (a) The backscattered electron image of a single crystal Cr$_{1/4}$NbSe$_2$. %
(b) The magnified view of the sample surface showing the positions where the chemical analyses were performed. %
 }
\end{figure}

\section*{Appendix:SEM-EDX measurements}

Figure \ref{SEM}(a) shows a backscattered electron image of a single crystal of Cr$_{1/4}$NbSe$_2$, which displays the spatial distribution the electron density. %
The uniform contrast of this image indicates that there is no significant inhomogeneity in chemical composition. %
Figure \ref{SEM}(b) shows an enlarged view on the sample surface. %
We focused on the 10 positions indicated by the cross marks, and performed chemical analyses by measuring characteristic X-rays emitted from the sample. %
The results are summarized in Table \ref{tab:SEM-EDX}, revealing that the ratios of the three elements is nearly independent of the positions. %
By normalizing the amount of Se to 8, the amounts of Cr and Nb are determined to be 0.74(4) and 3.98(8), respectively. %

\begin{table}[t]
\caption{\label{tab:SEM-EDX}
Results of SEM-EDX measurements.}
\renewcommand{\arraystretch}{1.3}
\begin{ruledtabular}
\begin{tabular}{lcccc}
Position & Cr & Nb & Se & Total(wt\%) \\
\hline
1 &3.95 &61.49 &35.34 &100.79 \\
2 &3.79 &61.16 &36.40 &101.34 \\
3 &3.70 &60.96 &36.11 &100.77 \\
4 &3.95 &60.89 &36.74 &101.58 \\
5 &3.71 &60.97 &35.77 &100.44 \\
6 &3.36 &60.89 &36.16 &100.41 \\
7 &3.92 &61.38 &35.88 &101.18 \\
8 &3.81 &60.90 &35.84 &100.55 \\
9 &3.38 &61.24 &35.41 &100.03 \\
10 &3.92 &61.61 &34.69 &100.22 \\
\hline
ave. & 0.74(4) &3.98(8) & 8 & \\
\end{tabular}
\end{ruledtabular}
\end{table}

\section*{Appendix: constant-energy cuts of the neutron inelastic scattering spectra}
\begin{figure}[b]
\includegraphics[keepaspectratio,width=0.95\linewidth]{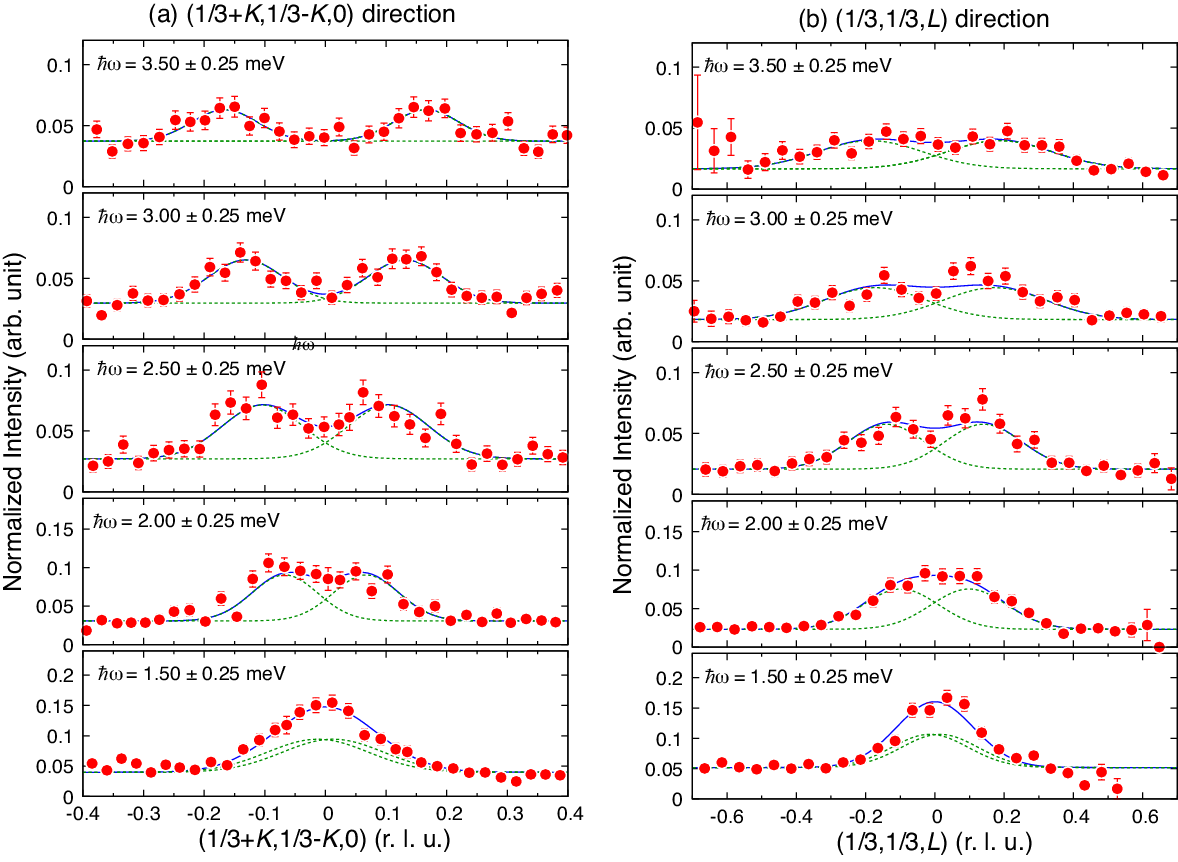}
\caption{\label{HRC_appendix} Constant-energy cuts of the excitation spectra along the (a) $(1/3+K,1/3-K,0)$ and (b) $(1/3,1/3,L)$ lines at different energies. Dotted and solid lines are Gaussian functions fitted to the data. %
 }
\end{figure}

Figures \ref{HRC_appendix}(a) and \ref{HRC_appendix}(b) show the constant-energy profiles along the $(1/3+K,1/3-K,0)$ and $(1/3,1/3,L)$ lines, which are extracted from the observed spectra shown in Figs. \ref{HRCdata}(h) and \ref{HRCdata}(i), respectively. %
For both directions, we observed two peaks moving away from each other as the energy transfer increases, being consistent with the Nambu-Goldstone modes emerging from the magnetic elastic Bragg peak at $(1/3,1/3,0)$. 

\clearpage

\bibliography{CrNb4Se8}

\end{document}